\documentclass{article}
\usepackage{spconf,amsmath,amssymb,graphicx,hyperref}
\usepackage{booktabs,multirow}
\usepackage{enumitem}
\setlist{topsep=2pt,itemsep=1pt,parsep=0pt,leftmargin=1.3em}
\makeatletter
\renewcommand\section{\@startsection{section}{1}{\z@}{2.1ex plus .6ex minus .2ex}{1.2ex plus .1ex}{\normalfont\Large\bfseries}}
\renewcommand\subsection{\@startsection{subsection}{2}{\z@}{1.7ex plus .5ex minus .2ex}{0.8ex plus .1ex}{\normalfont\large\bfseries}}
\renewcommand\subsubsection{\@startsection{subsubsection}{3}{\z@}{1.5ex plus .4ex minus .2ex}{0.7ex plus .1ex}{\normalfont\normalsize\bfseries}}
\makeatother
\usepackage{url}
\graphicspath{{figs/}}

\newcommand{\bench}{\textsc{KyaraBench}}
\newcommand{\model}{\textsc{KyaraEmbed}}

\title{CHARACTER IDENTITY IS NOT SPEAKER IDENTITY: \\KYARABENCH AND KYARAEMBED FOR CHARACTER VERIFICATION}
\name{Joonyong Park, Jerry Li}
\address{Spellbrush, CA, USA\\
\texttt{\small jyjoon97@gmail.com, jerry@sizigistudios.com}}

\begin{document}
\ninept
\maketitle

\begin{abstract}
A dubbed character keeps its identity while the voice actor changes, so character identity and speaker identity are distinct properties of one recording, yet speaker verification measures only the latter.
To address this gap, we propose \bench{}, a benchmark that scores character voice directly instead of speaker voice, built from 85 human-audited identities in a dubbed anime corpus.
It poses two challenging conditions: one that swaps the performer under a fixed character, and one that fixes the performer under changing characters.
Listening studies with 78 participants provide human reference scores for cross-performer verification and same-actor discrimination.
Speaker-verification baselines show increased errors under these character-specific conditions.
We then train \model{}, a compact encoder using multilingual character supervision, same-actor negatives, and a language-alignment term.
The model achieves the best performance on all character-specific conditions in the main comparison.
We release the benchmark, protocol, and encoder publicly.
\end{abstract}


\section{Introduction}
\label{sec:intro}

Voiced media gives characters a vocal identity beyond any single performer.
Audiences perceive several performers' voices --- recast or dubbed --- as one character's voice although no acoustic speaker identity is shared, and one performer's voice as several distinct characters although every cue points to one speaker \cite{lavan2019flexible}.
Systems that produce or manage such media inherit the corresponding question: for dubbing quality control and character voice design \cite{obin2016similarity,gresse2019similarity}, they must decide whether two voices belong to the same \emph{character}, not the same speaker.
However, existing evaluations cannot pose this question: speaker recognition takes the person as the unit of identity \cite{nagrani2017voxceleb,chung2018voxceleb2,fan2020cnceleb}.
The gap leaves both failure modes unmeasured: a system may split one dubbed character into two identities or merge two characters who share an actor.

We therefore propose \textsc{KyaraBench}, a benchmark built from the aligned dub pairs of a voiced-media corpus \cite{cai2024anim400k} through character identity clustering, naming from visual evidence, automatic quality gates, and human listening verification.
\bench{} offers five measures of character identity: same-language and cross-dub verification, same-actor discrimination, a pitch-matched variant, and cross-dub retrieval.
Every measure beyond the first is one that person-level protocols cannot make.
Listeners make more errors on the cross-dub task than on the single-language baseline, and accept a same-actor impostor more often than a pitch-matched control --- the same two confusions the benchmark's conditions isolate. 
These results provide a human reference for both conditions.

Furthermore, for the task the benchmark defines, we train \textsc{KyaraEmbed}, a single compact audio-only encoder with the same interface as a speaker embedding.
The model uses three datasets to provide cross-performer positives, same-actor impostors, and same-speaker cross-language pairs.
\textsc{KyaraEmbed} posts the best score among the systems compared in Table~\ref{tab:main} on all character-specific conditions.

Our contributions are 1) \textsc{KyaraBench}, the first benchmark of character identity under performer change, with audited labels, controlled impostor conditions, a protocol evaluated in listening studies, and a human reference on identical items, and 2) \textsc{KyaraEmbed}, a released encoder that verifies characters across performers better than every evaluated baseline.
The benchmark manifest, the evaluation code, and the model are publicly available\footnote{\url{github.com/sizigi/kyarabench}}.

\begin{figure}[t]
\centering
\includegraphics[width=\linewidth]{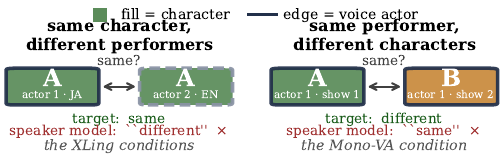}
\caption{Two cases of the speaker--character confusion. A dub changes the performer under a fixed character (left: answer \emph{same}), while one actor voices different characters in two shows (right: answer \emph{different}).}
\label{fig:task}
\end{figure}

\vspace{-1mm}
\section{Related Work}
\label{sec:related}

\textbf{Speaker verification benchmarks measure the performer.}
VoxCeleb and VoxCeleb2 established large-scale in-the-wild speaker verification \cite{nagrani2017voxceleb,chung2018voxceleb2}.
CN-Celeb extended it across genres including acted speech \cite{fan2020cnceleb}.
VoxMovies measured how far verification degrades when actors play parts \cite{brown2021playing}.
However, all of these keep the person as the unit of identity.
No protocol asks whether a system tracks the \emph{character} when the performer changes, or whether it separates two characters played by one performer.

\textbf{Character voice resources lack verified identity.}
Anim-400K aligns Japanese--English dubbed segments at scale, but it diarises speakers per episode and names no characters \cite{cai2024anim400k}.
V2C collects animated-movie speech with character labels for visual voice cloning, but it carries no cross-performer protocol \cite{chen2022v2c}.
Automatic voice casting shows that perceived similarity between acted voices is measurable, but it provides no controlled impostors \cite{obin2016similarity,gresse2019similarity}.
Consequently, neither direction of the speaker--character confusion can be scored on any existing resource.

\textbf{Benchmark methodology.}
SUPERB and SUPERB-SG pair each claimed capability with a protocol that isolates it \cite{yang2021superb,tsai2022superbsg}.
Cross-challenge listening studies set the methodology for human references \cite{cooper2021voices}.
Closer to our domain, there is a study comparing speaker-embedding distances against human similarity judgements \cite{kishi2026similarity}.
However, this evaluation discipline has not yet reached the speaker--character distinction: speaker similarity has metrics, while character identity does not.

\section{KyaraBench}
\label{sec:bench}

\begin{table}[t]
\centering
\caption{Construction funnel from Anim-400K to the released benchmark. Utts counts Japanese source utterances; Clips counts retained Japanese and English audio files.}
\label{tab:funnel}
\setlength{\tabcolsep}{4.5pt}
\begin{tabular}{l r r r}
\toprule
Stage & Identities & Utts & Clips \\
\midrule
Anim-400K aligned dub pairs & --- & 436{,}887 & 858{,}986 \\
show-level clustering & 9{,}399 & --- & --- \\
matched to a character portrait & 4{,}907 & --- & --- \\
$\geq$3 agreeing votes, cast-listed & 385 & --- & --- \\
coverage and quality gates & 114 & 1{,}828 & 3{,}656 \\
human annotation (released) & 85 & 986 & 1{,}616 \\
\bottomrule
\end{tabular}
\end{table}

\subsection{Construction}
\label{ssec:construction}
We build the benchmark from the Anim-400K corpus \cite{cai2024anim400k}, chosen because it holds both cross-lingual dubbing and multi-character acting by a single voice actor.
The corpus supplies aligned Japanese--English dub pairs together with the source video of every clip. 
However, its speaker labels are episode-local rather than show-wide, and no characters are named.

Table~\ref{tab:funnel} traces the funnel. Construction proceeds in an automatic filtering stage and a human annotation stage.
The automatic stage starts from the audio tracks.
Since broadcast mixes are noisy, every clip is first restored with Sidon \cite{nakata2026sidon}.
We then cluster episode-local character groups into show-level character identities, using embedding centroids from a publicly available speaker embedding model trained on an anime dataset\footnote{\url{huggingface.co/litagin/anime_speaker_embedding_ecapa_tdnn_groupnorm}}. 
During clustering, two different speaker labels from the same episode can never be linked, so groups merge only across episodes when their centroids are close. 

Next, we used visual evidence to name those identities.
We detect anime faces in the source video frames with an open YOLOv8 anime-face detector\footnote{\url{huggingface.co/deepghs/anime_face_detection}}.
Then, we match each crop against the show's public character portraits with CCIP\footnote{\url{huggingface.co/deepghs/ccip}}, a CLIP-style \cite{radford2021clip} character image encoder.
A cluster is approved when it matches a specific character at least three times with 75\% agreement or higher.

Next, we remove the outlier clips within each cluster.
A clip is removed when its Whisper \cite{radford2023whisper} transcription error rate exceeds $.3$, or when its UTMOS \cite{saeki2022utmos} score falls below $2.0$ for Japanese and $3.0$ for English clips.
It is also removed when the transcript, the pitch track, or the speaker embedder indicates two speakers inside the clip, or when its pitch sits more than five semitones from the identity's own register.
When the two-speaker evidence localises a clean change point, the clip is instead trimmed to its majority span.
Ninety-five released clips carry such a span annotation.
An identity survives only with at least six Japanese clips spanning at least two episodes.
The automatic filtering stage yields 114 identities and 1{,}828 Japanese--English clip pairs.

Finally, we run the human annotation stage, as an identity whose clips are one consistent voice passes every automatic test even when its name is wrong.
Annotators watched the source video clip beside the character's reference portrait, name, and show, and answered whether the speaking voice belongs to the pictured character.
A judgement was collected for all 1{,}828 of the clips (895 \emph{yes}, 91 \emph{off-screen but the voice matches}, 772 \emph{no}, 70 \emph{cannot tell}); keeping clips whose final judgement is \emph{yes} or \emph{off-screen}, leaves 85 identities and 986 clips.
The English sides pass the same gates on the dub track, and 630 of them are released.

\vspace{-1mm}
\subsubsection{Trial sampling}
\label{sssec:trials}
The evaluation trials are sampled from the surviving clips rather than enumerated exhaustively.
Each identity contributes at most twelve positive pairs per condition, with this cap applied separately to Japanese and English in \textsc{Mono}.
Impostor pairs are drawn at up to three per positive, under each condition's own constraint on who the impostor may be.
Every list is drawn once by a seeded random draw and shipped with the benchmark, so all systems are scored on identical trials.

\begin{table}[t]
\centering
\caption{Conditions of \textsc{KyaraBench} (positives / impostors). \textsc{Mono} pools the Japanese and English monolingual trials. \textsc{Mono-VA} adds impostors only and is scored against the \textsc{Mono} positives. Retrieval is a $53$-way gallery (chance R@1 $=.019$).}
\label{tab:cells}
\setlength{\tabcolsep}{5pt}
\begin{tabular}{l l r}
\toprule
Condition & Confusion tested & Trials (+/$-$) \\
\midrule
\textsc{Mono} & same char, same language & 1778 / 5327 \\
\textsc{XLing} & same char, dubbed performer & 646 / 1812 \\
\textsc{Mono-VA} & diff char, shared actor & --- / 309 \\
\textsc{XLing-K} & bundled, unconstrained impostor & 636 / 1908 \\
\textsc{XLing-KP} & bundled, pitch-matched impostor & 648 / 1944 \\
retrieval & Japanese query $\to$ English gallery & 424 queries \\
\bottomrule
\end{tabular}
\end{table}
\begin{table*}[t]
\centering
\caption{\textsc{KyaraBench} results (EER; lower is better, except R@1). Conditions as in Table~\ref{tab:cells}. $\dagger$: paired identity-bootstrap 95\% interval clear of zero against every speaker-verification baseline (ECAPA, ReDimNet, WavLM-SV). Best in bold.}
\label{tab:main}
\begin{tabular}{l l cccccc}
\toprule
 & System & \textsc{Mono} & \textsc{XLing} & \textsc{Mono-VA} & \textsc{XLing-K} & \textsc{XLing-KP} & R@1 \\
\midrule
shortcut & F0 statistics & .395 & .418 & .392 & .338 & .477 & .087 \\
\midrule
\multirow{4}{*}{zero-shot}
 & ECAPA-TDNN (VoxCeleb) & \textbf{.138} & .367 & .305 & .344 & .437 & .130 \\
 & ReDimNet-b2 (VoxCeleb2) & .156 & .383 & .295 & .316 & .418 & .068 \\
 & WavLM-base+ SV & .201 & .314 & .288 & .267 & .420 & .052 \\
 & Qwen3-Omni audio tower & .387 & .455 & .417 & .356 & .465 & .137 \\
\midrule
domain & ECAPA-TDNN (domain fine-tuned) & .149 & .271 & .252 & .225 & .344 & .210 \\
\midrule
ours & \model{} & .155 & \textbf{.215}$^\dagger$ & \textbf{.207}$^\dagger$ & \textbf{.159}$^\dagger$ & \textbf{.281}$^\dagger$ & \textbf{.325}$^\dagger$ \\
\bottomrule
\end{tabular}
\end{table*}
\vspace{-1mm}
\subsection{Task and conditions}
Character verification determines whether two audio bundles belong to the same \emph{character}, irrespective of language or voice actor.
Similarity is the cosine between the evaluated system's clip embeddings, mean-pooled over each bundle.
The correct answer is the human-audited character label of Section~\ref{ssec:construction}; the system hears only the audio. 
We build one deliberately hard condition for each point where speaker and character identity can be confused, with trial counts in Table~\ref{tab:cells} (Fig.~\ref{fig:task}).
Beyond the table: bundles hold four clips per side, pitch-matched impostors sit within two semitones of the target's median pitch, and the retrieval gallery holds the 53 characters with at least four English clips at the draw, each queried by eight four-clip Japanese bundles.
\vspace{-1mm}
\subsection{Protocol and validity}
A system under evaluation provides one function that turns an audio clip into an embedding vector.
The released loader feeds it every clip as mono 16\,kHz audio.
Each trial is then scored as the cosine between its two sides, a bundle represented by the mean of its members' embeddings.
Verification conditions are summarised by the equal-error rate (EER), the error at the threshold where false acceptances equal misses, so no operating threshold has to be chosen.
Retrieval is summarised by R@1 against the fixed gallery.
Uncertainty is reported by resampling identities rather than trials, because trials that share an identity are not independent.
Any free parameter of a scoring rule is fixed on conditions other than the one reported.
The human annotation stage roughly halves \textsc{Mono} error while leaving \textsc{XLing} within its interval.

\section{KyaraEmbed}
\label{sec:method}

Beyond measuring the two failure modes, we further build a model to resolve them.
\model{} is a single compact audio-only encoder (ECAPA-TDNN \cite{desplanques2020ecapa}, 192-d output), fine-tuned from the public anime-domain speaker embedder of Section~\ref{sec:bench}.

\textbf{Structuring the datasets.}
Within one language, the clips of a character are also the clips of one actor, so cross-performer positives do not exist monolingually.
An internal dataset supplies them: 86 characters, each recorded in four languages by four different actors, about 91 hours in total.
Same-voice impostors are vanishingly rare under random sampling, so we draw them deliberately.
Cast lists recover each anime character's Japanese actor, and cross-show pairs of characters sharing an actor are sampled as tied pairs, so every training batch contains the hard condition.
Finally, a change of language always brings a change of performer, so none of the corpora above isolates the language axis. 
The tri-jek corpus\footnote{\url{sites.google.com/site/shinnosuketakamichi/research-topics/tri-jek_corpus}} supplies the missing contrast: one trilingual speaker holding identity constant while the language changes.

\textbf{Objective and training design.}
The base objective is supervised contrastive learning \cite{khosla2020supcon} over character identities, with each character's clips spread across languages so the pull of the loss crosses performers by construction.
A margin $m^{+}{=}.05$ then demands extra similarity from exactly the cross-actor positives: $s_{ij} = \cos(z_i,z_j) - m^{+}\,\mathbb{1}[\text{same char, diff.\ actor}]$.
A margin $m^{-}{=}.10$ pushes the tied same-actor pairs apart: $s_{ij} = \cos(z_i,z_j) + m^{-}\,\mathbb{1}[\text{diff.\ char, same actor}]$.
The \emph{language-alignment} term adds eight pairs to every batch, each the same trilingual speaker in two different languages.
It is an InfoNCE loss \cite{oord2018cpc} (weight $\lambda{=}.3$) whose positive is that cross-language pair and whose negatives are the batch's character clips, held fixed: $\cos(z_a,z_b)$ must beat every $\cos(z_a,z_k)$.
This term encourages cross-language consistency for the same speaker while contrasting these pairs against character clips.
Training runs 1{,}500 steps of AdamW at $10^{-5}$, with batches of eight characters and four tied actor pairs at four clips each, at temperature $.07$.
The training and evaluation benchmark sets share no clips, episodes, shows, and character identities.

\section{Experiments}
\label{sec:exp}
\begin{table}[t]
\centering
\caption{Human reference on identical items, from two listening studies. Left: balanced error rate (BER) on 320 verification trials (48 listeners). Right: the share of impostor trials correctly rejected in the same-actor study (200 trials, 30 listeners), for impostors that share the voice actor (same-VA) and for pitch-matched impostors with a different actor (diff-VA); the penalty is the drop in that rejection rate when the actor is shared (diff-VA - same-VA).}
\label{tab:human}
\setlength{\tabcolsep}{2.0pt}
\begin{tabular}{l ccc ccc}
\toprule
& \multicolumn{3}{c}{BER $\downarrow$} & \multicolumn{3}{c}{impostor rejection $\uparrow$} \\
\cmidrule(lr){2-4}\cmidrule(lr){5-7}
 & \textsc{Mono} & \textsc{XLing} & K4 & same-VA & diff-VA & penalty \\
\midrule
Human & .226 & .318 & .319 & \textbf{.820} & \textbf{.907} & $+.087$ \\
ECAPA-TDNN & .143 & .375 & .375 & .620 & .820 & $+.200$ \\
domain FT & \textbf{.107} & .232 & .292 & .720 & .820 & $+.100$ \\
\model{} & .143 & \textbf{.214} & \textbf{.229} & .760 & .780 & $\mathbf{+.020}$ \\
\bottomrule
\end{tabular}
\end{table}
\begin{table*}[t]
\centering
\caption{Ablation ladder from the domain fine-tuned model to \model{} on \bench{} (conditions as in Table~\ref{tab:cells}; best in bold). The first two rungs change the training data; every rung below the second rule keeps the full corpus, the initialisation from the plain-SupCon rung, and the 1{,}500-step schedule, toggling one ingredient at a time.}
\label{tab:ablation}
\begin{tabular}{l cccccc}
\toprule
Configuration & \textsc{Mono} & \textsc{XLing} & \textsc{Mono-VA} & \textsc{XLing-K} & \textsc{XLing-KP} & R@1 \\
\midrule
domain fine-tuned & .149 & .271 & .252 & .225 & .344 & .210 \\
+ internal-dataset SupCon (adapted) & .196 & .258 & .281 & .190 & .323 & .226 \\
+ anime corpus, plain SupCon & .154 & .234 & .249 & .165 & .287 & .321 \\
\midrule
+ actor-tied sampling, no margins & .134 & .266 & .239 & .186 & .307 & .262 \\
+ margins, untied characters & \textbf{.133} & .263 & .246 & .181 & .306 & .274 \\
+ tie and margins ($\lambda{=}0$) & .136 & .269 & .243 & .186 & .303 & .264 \\
\midrule
+ language alignment, $\lambda{=}.3$ (\model{}) & .155 & \textbf{.215} & .207 & .159 & \textbf{.281} & \textbf{.325} \\
\hphantom{+ language alignment, }$\lambda{=}1.0$ & .151 & \textbf{.215} & \textbf{.205} & \textbf{.156} & .282 & .297 \\
\bottomrule
\end{tabular}
\end{table*}

\subsection{Setup}

\textbf{Baselines.}
We evaluate our model against several \emph{zero-shot} baselines using publicly available, off-the-shelf checkpoints. These include ECAPA-TDNN \cite{desplanques2020ecapa} and ReDimNet-b2 \cite{yakovlev2024redimnet} (both trained on VoxCeleb), the WavLM-base+ \cite{chen2022wavlm} speaker-verification head, and the audio tower of Qwen3-Omni \cite{qwen2025omni} (mean-pooled to extract embeddings). Additionally, we include the ECAPA-TDNN embedder fine-tuned with 10k hours of anime speech as a domain-matched anchor; it also serves as the initialization point for our own model.

\textbf{Human Listening Studies.}
To establish a human reference point, we conducted two distinct listening studies using trial subsets drawn directly from the machine evaluation lists. In the first study, 48 native Japanese listeners evaluated the monolingual and cross-lingual conditions, ensuring three independent annotators rated each item. A separate cohort of 30 listeners evaluated the more challenging same-actor condition. 
Each participant rated twenty randomly mixed trials, judging whether the voices could belong to the same fictional character regardless of language or voice actor, and reporting confidence on a five-point scale.
To ensure that prior knowledge did not bias the evaluation, we verified that self-reported show recognition remained below 7\% across all ratings.
Table~\ref{tab:human} reports balanced error rate (BER), defined as the mean of false-acceptance and false-rejection rates, from the pooled human binary decisions.
Model BER is computed on the same items at each model's cosine-score EER threshold.

\subsection{Main results}
As shown in Table~\ref{tab:main}, the speaker-verification baselines perform reasonably well on the \textsc{Mono} condition, where the voice actor remains the same.
However, their error rates increase under the two character-specific scenarios. 
First, when a character's performer changes across a dub (\textsc{XLing}), their performance drops drastically; ECAPA's error nearly triples, falling behind even human listeners on the bundled items (Table~\ref{tab:human}). 
On \textsc{Mono-VA}, the three speaker-verification baselines yield EERs between $.288$ and $.305$.
We also confirm this failure is not merely a matter of pitch shift, as the F0 baseline performs at chance level on the pitch-matched \textsc{XLing-KP} condition. 
\model{}, by contrast, successfully isolates character identity, by achieving the best performance among the systems in Table~\ref{tab:main} on \textsc{XLing}, \textsc{Mono-VA}, both bundled conditions, and retrieval.
\model{} concedes only the \textsc{Mono} condition to ECAPA, which is expected as \textsc{Mono} purely tests single-actor speaker verification rather than cross-performer character identity.

Single-clip \textsc{XLing} positives exclude aligned Japanese--English segment pairs, whereas approximately 79\% of positives in each bundled condition contain at least one such pair.
With the same impostors, \model{} yields \textsc{XLing-K} EERs of $.173$ and $.155$ on positive subsets without and with such pairs, respectively.
Replacing the standard impostors with 397 within-show impostors from 16 shows, while retaining the same positives, changes its \textsc{XLing} EER from $.215$ to $.201$.

\vspace{-1mm}
\subsection{The human reference}
Human listeners struggle with the exact confusions the benchmark is designed to measure, validating the difficulty of the task (Table~\ref{tab:human}). 
When evaluating characters across a dub (\textsc{XLing}), human error increases significantly, showing a paired gap of $+.092$ $[+.010, +.173]$ compared to the monolingual baseline. 
On the human-rated subset, bundling does not improve BER for either listeners or the evaluated models.
Furthermore, listeners are easily misled when a single actor voices multiple characters (\textsc{Mono-VA}). 
Humans pay a significant penalty on same-actor impostors (actor bootstrap $[+.001,+.157]$), particularly when the pitch is similar.
\model{} shows a smaller rejection-rate gap between the two impostor conditions than humans ($.020$ vs.\ $.087$; Table~\ref{tab:human}).
However, its absolute rejection rates are lower in both conditions (same-VA: $.760$ vs.\ $.820$; diff-VA: $.780$ vs.\ $.907$).
These results provide a human reference for listeners with limited familiarity with the shows.

Human and machine difficulty also align at the item level: the correlation between our model's per-item correctness and the three raters' accuracy is clear of zero on every rated condition (item bootstrap; $\phi$ $.29$ monolingual, $.26$ and $.25$ cross-lingual), whereas the domain fine-tuned model aligns only monolingually ($.26$) and shows correlations closer to zero across the dub ($.08$ and $.00$).

\begin{figure}[t]
\centering
\includegraphics[width=0.96\linewidth]{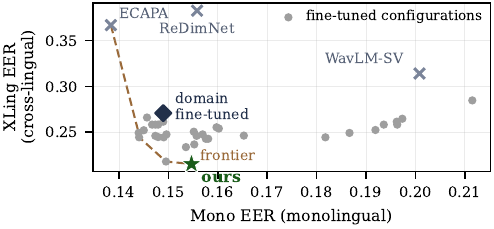}
\caption{\textsc{Mono} vs.\ \textsc{XLing} EER for 31 fine-tuned configurations from our ablations (grey), the domain fine-tuned model, the speaker-verification baselines, and our model. The dashed line shows the empirical Pareto frontier among the configurations plotted.}
\label{fig:frontier}
\end{figure}

\vspace{-1mm}
\subsection{Ablation studies: what buys the capability}
Table~\ref{tab:ablation} walks from the domain fine-tuned model to \model{} one ingredient at a time.
The data rungs come first: internal-dataset character supervision alone, the \emph{adapted} rung, opens the cross-lingual axis only modestly and sacrifices \textsc{Mono} performance to do it, while adding the anime corpus under plain supervised contrastive learning already yields most of the bundled and retrieval gains.
The actor-conditional rungs then trade performance between the axes rather than advancing both simultaneously: with the language term off, tied sampling and margins in any combination yield the best \textsc{Mono} results but surrender most of the cross-lingual gain, with the three variants sitting within a few thousandths of one another.
The language-alignment term resolves this trade-off into the final model: at $\lambda{=}.3$, every character condition improves cross-lingually at a small \textsc{Mono} cost. A paired bootstrap confirms this pattern: the $\lambda{=}0$ family is significantly better on \textsc{Mono} and significantly worse on \textsc{XLing} than \model{}.
The same-actor condition falls along the same path, performing worst under the adapted rung and best under the full recipe, though its wide intervals make this an ordering rather than a definitive margin.

The evaluated configurations exhibit different balances between \textsc{Mono} and \textsc{XLing} EER (Fig.~\ref{fig:frontier}).
We selected \model{} based on its joint ranking on \textsc{XLing}, \textsc{Mono-VA}, and R@1 on \bench{}.
It achieves the best \textsc{XLing} performance while incurring only a minimal decrease in \textsc{Mono} performance compared to the strict monolingual baselines.

\section{Conclusion}
\label{sec:conclusion}
Character identity survives a change of performer; speaker identity does not.
\bench{} operationalizes this distinction through verification and retrieval tasks, the listening studies provide human reference scores, and \model{} outperforms the baselines in Table~\ref{tab:main} on the character-specific conditions.
The study is limited in scope to acted Japanese anime speech and its English dubs.
Future work will explore representations that jointly support speaker and character verification.

\vfill\pagebreak

\section{Acknowledgments}
\textbf{Ethics statement.}
The listening study involved non-invasive perceptual judgments of speech. 
Participants were recruited through a crowdsourcing platform and personal contacts, with crowdsourced participants receiving monetary compensation. 
Participation was voluntary, and participants were informed beforehand that their responses would be used for research and reported only in aggregate. 

We redistributes no source audio or video content. Source content rights remain with their respective rights holders; we neither claim ownership nor relicense the underlying media.

\textbf{Use of Generative AI Disclosure.} Claude (Anthropic) was used to assist with language revision of the manuscript and experimental code development. All AI-generated outputs were reviewed and verified by the authors.

\bibliographystyle{IEEEbib}
\bibliography{strings,bib/refs}

@inproceedings{desplanques2020ecapa,
  author    = {Brecht Desplanques and Jenthe Thienpondt and Kris Demuynck},
  title     = {{ECAPA-TDNN}: Emphasized Channel Attention, Propagation and Aggregation in {TDNN} Based Speaker Verification},
  booktitle = {Proc. Interspeech},
  year      = {2020}
}

@article{chen2022wavlm,
  author  = {Sanyuan Chen and Chengyi Wang and Zhengyang Chen and Yu Wu and Shujie Liu and Zhuo Chen and Jinyu Li and Naoyuki Kanda and Takuya Yoshioka and Xiong Xiao and Jian Wu and Long Zhou and Shuo Ren and Yanmin Qian and Yao Qian and Jian Wu and Michael Zeng and Xiangzhan Yu and Furu Wei},
  title   = {{WavLM}: Large-Scale Self-Supervised Pre-Training for Full Stack Speech Processing},
  journal = {IEEE Journal of Selected Topics in Signal Processing},
  volume  = {16},
  number  = {6},
  pages   = {1505--1518},
  year    = {2022}
}

@inproceedings{yakovlev2024redimnet,
  author    = {Ivan Yakovlev and Rostislav Makarov and Andrei Balykin and Pavel Malov and Anton Okhotnikov and Nikita Torgashov},
  title     = {Reshape Dimensions Network for Speaker Recognition},
  booktitle = {Proc. Interspeech},
  year      = {2024}
}

@inproceedings{nagrani2017voxceleb,
  author    = {Arsha Nagrani and Joon Son Chung and Andrew Zisserman},
  title     = {{VoxCeleb}: A Large-Scale Speaker Identification Dataset},
  booktitle = {Proc. Interspeech},
  year      = {2017}
}

@inproceedings{khosla2020supcon,
  author    = {Prannay Khosla and Piotr Teterwak and Chen Wang and Aaron Sarna and Yonglong Tian and Phillip Isola and Aaron Maschinot and Ce Liu and Dilip Krishnan},
  title     = {Supervised Contrastive Learning},
  booktitle = {Advances in Neural Information Processing Systems},
  year      = {2020}
}

@inproceedings{radford2023whisper,
  author    = {Alec Radford and Jong Wook Kim and Tao Xu and Greg Brockman and Christine McLeavey and Ilya Sutskever},
  title     = {Robust Speech Recognition via Large-Scale Weak Supervision},
  booktitle = {Proc. ICML},
  year      = {2023}
}

@inproceedings{cai2024anim400k,
  author    = {Kevin Cai and Chonghua Liu and David M. Chan},
  title     = {{Anim-400K}: A Large-Scale Dataset for Automated End-to-End Dubbing of Video},
  booktitle = {Proc. IEEE ICASSP},
  year      = {2024}
}

@inproceedings{brown2021playing,
  title={Playing a part: Speaker verification at the movies},
  author={Brown, Andrew and Huh, Jaesung and Nagrani, Arsha and Chung, Joon Son and Zisserman, Andrew},
  booktitle={IEEE International Conference on Acoustics, Speech and Signal Processing (ICASSP)},
  pages={6174--6178},
  year={2021}
}

@inproceedings{chen2022v2c,
  title={{V2C}: Visual voice cloning},
  author={Chen, Qi and Tan, Mingkui and Qi, Yuankai and Zhou, Jiaqiu and Li, Yuanqing and Wu, Qi},
  booktitle={IEEE/CVF Conference on Computer Vision and Pattern Recognition (CVPR)},
  pages={21242--21251},
  year={2022}
}

@article{obin2016similarity,
  title={Similarity search of acted voices for automatic voice casting},
  author={Obin, Nicolas and Roebel, Axel},
  journal={IEEE/ACM Transactions on Audio, Speech, and Language Processing},
  volume={24},
  number={9},
  pages={1642--1651},
  year={2016}
}

@inproceedings{gresse2019similarity,
  title={Similarity metric based on {S}iamese neural networks for voice casting},
  author={Gresse, Adrien and Quillot, Mathias and Dufour, Richard and Labatut, Vincent and Bonastre, Jean-Fran{\c{c}}ois},
  booktitle={IEEE International Conference on Acoustics, Speech and Signal Processing (ICASSP)},
  pages={6585--6589},
  year={2019}
}

@inproceedings{chung2018voxceleb2,
  author    = {Joon Son Chung and Arsha Nagrani and Andrew Zisserman},
  title     = {{VoxCeleb2}: Deep Speaker Recognition},
  booktitle = {Proc. Interspeech},
  year      = {2018}
}

@inproceedings{fan2020cnceleb,
  author    = {Yue Fan and Jiawen Kang and Lantian Li and Kaicheng Li and Haolin Chen and Sitong Cheng and Pengyuan Zhang and Ziya Zhou and Yunqi Cai and Dong Wang},
  title     = {{CN-Celeb}: A Challenging {C}hinese Speaker Recognition Dataset},
  booktitle = {Proc. IEEE ICASSP},
  year      = {2020}
}

@inproceedings{yang2021superb,
  author    = {Shu-wen Yang and Po-Han Chi and Yung-Sung Chuang and Cheng-I Jeff Lai and Kushal Lakhotia and Yist Y. Lin and Andy T. Liu and Jiatong Shi and Xuankai Chang and Guan-Ting Lin and Tzu-Hsien Huang and Wei-Cheng Tseng and Ko-tik Lee and Da-Rong Liu and Zili Huang and Shuyan Dong and Shang-Wen Li and Shinji Watanabe and Abdelrahman Mohamed and Hung-yi Lee},
  title     = {{SUPERB}: Speech Processing Universal {PERformance} Benchmark},
  booktitle = {Proc. Interspeech},
  year      = {2021}
}

@inproceedings{tsai2022superbsg,
  author    = {Hsiang-Sheng Tsai and Heng-Jui Chang and Wen-Chin Huang and Zili Huang and Kushal Lakhotia and Shu-wen Yang and Shuyan Dong and Andy T. Liu and Cheng-I Jeff Lai and Jiatong Shi and Xuankai Chang and Phil Hall and Hsuan-Jui Chen and Shang-Wen Li and Shinji Watanabe and Abdelrahman Mohamed and Hung-yi Lee},
  title     = {{SUPERB-SG}: Enhanced Speech Processing Universal {PERformance} Benchmark for Semantic and Generative Capabilities},
  booktitle = {Proc. ACL},
  pages     = {8479--8492},
  year      = {2022}
}

@inproceedings{cooper2021voices,
  author    = {Erica Cooper and Junichi Yamagishi},
  title     = {How do Voices from Past Speech Synthesis Challenges Compare Today?},
  booktitle = {Proc. ISCA Speech Synthesis Workshop (SSW)},
  year      = {2021}
}

@inproceedings{saeki2022utmos,
  author    = {Takaaki Saeki and Detai Xin and Wataru Nakata and Tomoki Koriyama and Shinnosuke Takamichi and Hiroshi Saruwatari},
  title     = {{UTMOS}: {UT}okyo-{S}aru{L}ab System for {VoiceMOS} Challenge 2022},
  booktitle = {Proc. Interspeech},
  year      = {2022}
}

@inproceedings{nakata2026sidon,
  author    = {Wataru Nakata and Yuki Saito and Yota Ueda and Hiroshi Saruwatari},
  title     = {Sidon: Fast and Robust Open-Source Multilingual Speech Restoration for Large-scale Dataset Cleansing},
  booktitle = {Proc. IEEE ICASSP},
  year      = {2026}
}

@article{qwen2025omni,
  author  = {Jin Xu and others},
  title   = {{Qwen3-Omni} Technical Report},
  journal = {arXiv preprint arXiv:2509.17765},
  year    = {2025}
}

@inproceedings{kishi2026similarity,
  author    = {Minoru Kishi and Hayato Yagi and Shinnosuke Takamichi and Yuki Saito},
  title     = {Do speech foundation models perceive speaker similarity as humans do?},
  booktitle = {Proc. Interspeech},
  year      = {2026}
}

@article{lavan2019flexible,
  author  = {Nadine Lavan and A. Mike Burton and Sophie K. Scott and Carolyn McGettigan},
  title   = {Flexible voices: Identity perception from variable vocal signals},
  journal = {Psychonomic Bulletin \& Review},
  volume  = {26},
  number  = {1},
  pages   = {90--102},
  year    = {2019}
}

@inproceedings{radford2021clip,
  author    = {Alec Radford and Jong Wook Kim and Chris Hallacy and Aditya Ramesh and Gabriel Goh and Sandhini Agarwal and Girish Sastry and Amanda Askell and Pamela Mishkin and Jack Clark and Gretchen Krueger and Ilya Sutskever},
  title     = {Learning Transferable Visual Models From Natural Language Supervision},
  booktitle = {Proc. ICML},
  year      = {2021}
}

@article{oord2018cpc,
  author  = {A{\"a}ron van den Oord and Yazhe Li and Oriol Vinyals},
  title   = {Representation Learning with Contrastive Predictive Coding},
  journal = {arXiv preprint arXiv:1807.03748},
  year    = {2018}
}

\end{document}